\documentstyle[12pt,epsf]{article}
\def\fbi{\rm fb^{-1}}
\def\tb{\bar t}

\def\bb{\bar b}

\def\lsim{\mathrel{\raise.3ex\hbox{$<$\kern-.75em\lower1ex\hbox{$\sim$}}}}
\def\gsim{\mathrel{\raise.3ex\hbox{$>$\kern-.75em\lower1ex\hbox{$\sim$}}}}
\newcommand{ \slashchar }[1]{\setbox0=\hbox{$#1$}   
   \dimen0=\wd0                                     
   \setbox1=\hbox{/} \dimen1=\wd1                   
   \ifdim\dimen0>\dimen1                            
      \rlap{\hbox to \dimen0{\hfil/\hfil}}          
      #1                                            
   \else                                            
      \rlap{\hbox to \dimen1{\hfil$#1$\hfil}}       
      /                                             
   \fi}                                             %

\def\gev{\,{\rm GeV}}

\def\to{\rightarrow}

\def\be{\begin{equation}}
\def\ee{\end{equation}}
\def\bea{\begin{eqnarray}}
\def\eea{\end{eqnarray}}
\def\atversim#1#2{\lower0.7ex\vbox{\baselineskip\zatskip\lineskip\zatskip
  \lineskiplimit 0pt\ialign{$\matth#1\hfil##\hfil$\crcr#2\crcr\sim\crcr}}}

\renewcommand{\thefootnote}{\fnsymbol{footnote}}

\newcounter{appendixc}
\newcounter{subappendixc}[appendixc]
\newcounter{subsubappendixc}[subappendixc]

\renewcommand{\appendix}[1] {\vspace*{0.6cm}
        \refstepcounter{appendixc}
        \setcounter{figure}{0}
        \setcounter{table}{0}
        \setcounter{equation}{0}
        \renewcommand{\thefigure}{\Alph{appendixc}.\arabic{figure}}
        \renewcommand{\thetable}{\Alph{appendixc}.\arabic{table}}
        \renewcommand{\theappendixc}{\Alph{appendixc}}
        \renewcommand{\theequation}{\Alph{appendixc}.\arabic{equation}}
        \noindent{\bf Appendix \theappendixc #1}\par\vspace*{0.4cm}}

\begin{document}

\begin{titlepage}
\rightline{\vbox{\halign{&#\hfil\cr
&MADPH-01-1213\cr
&hep-ph/0106344\cr
&July 2001\cr}}}
\begin{center}

{\Large\bf  Top-quark spin correlation at Linear Colliders\\
with anomalous couplings}

\bigskip

\normalsize
{\bf  Z.-H. Lin$^1$, T. Han$^2$, T. Huang$^1$,
J.-X. Wang$^1$ and X. Zhang$^1$ } \\
\vskip .3cm
$^1$Institute of High Energy Physics, Academia Sinica,\\
Beijing, 100039, P. R. China,\\
$^2$Department of Physics, University of Wisconsin, Madison, WI 53706, USA\\
\vskip .3in

\end{center}

\begin{abstract}

 We investigate the feasibility of probing anomalous
 top-quark couplings of $Wtb$, $Z t \bar{t}$, and $\gamma t \bar{t}$
 in terms of an effective Lagrangian with dimension-six operators
 at future $e^+e^-$ linear colliders with a c.~m.~energy
$\sqrt s \sim 500-800$ GeV.
 We first examine the constraints on these anomalous couplings
 from the $Z\to b \bar{b}$ data at LEP I
 and from unitarity considerations. We then consider in detail
the effects of anomalous couplings on $t \bar{t}$ spin correlations in the
top-pair production and decay with three spin bases:
the helicity, beamline and off-diagonal bases.
Our results show that the polarized beams are more suitable
for exploring the effects of different new operators.
For polarized beams, the helicity basis yields the best sensitivity.
\end{abstract}

\renewcommand{\thefootnote}{\arabic{footnote}}
\end{titlepage}


\section{Introduction}

 The top-quark physics is one of the most important
 topics at a next generation linear collider. Near and above the
threshold of a top quark and anti-top quark ($t\bar t$),
 there will be about $500-1000\ t\bar t$ pairs produced per one
$\fbi$ of integrated luminosity. Thus a high luminosity
linear collider may allow for a good determination \cite{toplc}
of the top-quark mass ($m_t$), its total width, the strong coupling
constant $\alpha_s$, and even new physics effects such as
from a virtual Higgs boson. Indeed, as the heaviest particle
observed so far with a mass at the electroweak scale,
the top quark may hold the key to new physics
in the electroweak sector \cite{peccei}.

Without knowing the underlying dynamics beyond the standard model
(SM), it is appropriate to parameterize new physics effects
at low energies by an effective Lagrangian
 \begin{equation}
 \label{eff}
 {\cal L}_{eff}={\cal L}_0+\frac{1}{\Lambda^2}\sum_i C_i O_i
                          +{\cal O}(\frac{1}{\Lambda^4}),
 \end{equation}
where $\Lambda$ is a cutoff scale above which new physics sets
in, and ${\cal L}_0$ is the SM Lagrangian,
$O_i$ the SM-gauge-invariant dimension-six operators.
$C_i$ represent the coupling strengths of $O_i$~\cite{linear}.
In this paper we study the possibility of probing for anomalous top-quark
couplings $\gamma t \tb$, $Z t \tb$ and $Wtb$ at Next Linear Colliders.

It has been shown in the literature that
angular distribution in
top-quark events at $e^+e^-$ colliders bring useful information on
the top-quark couplings, with which one can
constrain its
deviations from the SM~\cite{recent,talk}.
At $e^+e^-$ colliders the top-quark pair
is produced in a unique spin configuration, and the
electroweak decay products of polarized top quark
are strongly correlated to the spin axis.
Parke and Shadmi~\cite{parke} suggested several spin basis
(helicity, beamline and off-diagonal
basis) to discuss the $t\bar t$ spin correlation. On the other hand,
deviations from the SM may be observable via the top-quark spin
correlations which depend
sensitively on the couplings $\gamma t \tb$, $Z t \tb$ and $Wtb$.
The purpose of the current study is to explore
which spin basis described in Ref.~\cite{parke} is more suitable
for studying a given anomalous top-quark coupling than others.

This paper is organized as follows.
In section 2, we briefly review on effective
lagrangian and present explicitly the
dimension-6 operators which describe the  anomalous couplings $\gamma t \tb$,
$Z t \tb$ and $Wtb$, then we examine
the constraints on the coefficients
of these operators. In section 3, we present a detail calculation
of the top-quark spin correlation
, and finally
section 4 is
summary and conclusion.

\section{Effective interactions with top-quark}

In the linearly realized effective Lagrangian \cite{linear}, the
anomalous top-quark couplings are described by higher
dimensional operators.
Recently the dimension-six operators involving the top quark and
invariant under the SM gauge group were
reclassified and some are analyzed in Refs.~\cite{Whisnant,Renard}.
Here we list all eleven
dimension-six CP-even operators in Refs.~\cite{Whisnant,Renard}
which generate anomalous
couplings of $Z,\gamma,W^\pm$ to the top quark beyond the
SM interactions,
\bea \label{operators}
O_{\Phi q}^{(1)} &=&i\left [\Phi^{\dagger}D_{\mu}\Phi
      -(D_{\mu}\Phi)^{\dagger}\Phi\right ]\bar q_L \gamma^{\mu}q_L,\nonumber\\
O_{\Phi q}^{(3)}&=&i\left [\Phi^{\dagger}\tau^I D_{\mu}\Phi
        -(D_{\mu}\Phi)^{\dagger}\tau^I\Phi\right ]\bar q_L \gamma^{\mu}\tau^I
       q_L,\nonumber\\
O_{Db} &=&(\bar q_L D_{\mu} b_R) D^{\mu}\Phi
         +(D^{\mu}\Phi)^{\dagger}(\overline{D_{\mu}b_R}q_L),\nonumber\\
O_{bW\Phi} &=&\left [(\bar q_L \sigma^{\mu\nu}\tau^I b_R) \Phi
         +\Phi^{\dagger}(\bar b_R \sigma^{\mu\nu}\tau^I q_L)\right ]
          W^I_{\mu\nu},\nonumber\\
O_{qB} &=&\left [\bar q_L \gamma^{\mu} D^{\nu}q_L
         +\overline{D^{\nu}q_L} \gamma^{\mu} q_L\right ]
          B_{\mu\nu},\nonumber\\
O_{qW} &=&\left [\bar q_L \gamma^{\mu}\tau^I D^{\nu}q_L
         +\overline{D^{\nu}q_L} \gamma^{\mu}\tau^I q_L\right ]
          W^I_{\mu\nu},\nonumber\\
O_{t2} &=&i\left [\Phi^{\dagger}D_{\mu}\Phi
         -(D_{\mu}\Phi)^{\dagger}\Phi\right ]\bar t_R \gamma^{\mu}t_R,
\nonumber\\
O_{Dt} &=&(\bar q_L D_{\mu} t_R) D^{\mu}\widetilde\Phi
         +(D^{\mu}\widetilde\Phi)^{\dagger}(\overline{D_{\mu}t_R}q_L),
\nonumber\\
O_{tB\Phi} &=&\left [(\bar q_L \sigma^{\mu\nu} t_R) \widetilde\Phi
         +\widetilde\Phi^{\dagger}(\bar t_R \sigma^{\mu\nu} q_L)\right ]
          B_{\mu\nu},\nonumber\\
O_{tW\Phi} &=&\left [(\bar q_L \sigma^{\mu\nu}\tau^I t_R) \widetilde\Phi
         +\widetilde\Phi^{\dagger}(\bar t_R \sigma^{\mu\nu}\tau^I q_L)\right ]
          W^I_{\mu\nu},\nonumber\\
O_{tB} &=&\left [\bar t_R \gamma^{\mu} D^{\nu}t_R
         +\overline{D^{\nu}t_R} \gamma^{\mu} t_R\right ]
          B_{\mu\nu},
\eea
where $\Phi$ is the Higgs doublet,
$\widetilde\Phi = i\tau^2 \Phi^*$, $\bar q_L=(\bar t_L, \bar b_L)$
and $\tau^I$ are Pauli matrices.
In Eq.(2), some of the operators induce energy-dependent couplings,
some do not. If an anomalous coupling is function of energy, its effects
 on the physical quantity at different energy scale
 will be enhanced at high energy. In Table~\ref{one} we show explicitly the
energy dependence of various couplings.

\begin{table}[thb]
\begin{center}
\begin{tabular}{lcccc}\hline\hline
  &$~Zb\bar b~$ &$~Wt\bar b~$  & $~Zt\bar t~$ & $~\gamma t\bar t~$\\\hline
   SM          &1&1&1&1 \\
$O_{\Phi q}^{(1)}$&$1$&&$1$&\\
$O_{\Phi q}^{(3)}$&$1$&$1$&$1$&\\
$O_{Db}$&$E/v$&$E/v$&&\\
$O_{bW\Phi}$&$E/v$&$E/v$&&\\
$O_{qB}$&$E^2/v^2$&&$E^2/v^2$&$E^2/v^2$\\
$O_{qW}$&$E^2/v^2$&$E^2/v^2$&$E^2/v^2$&$E^2/v^2$\\
$O_{t2}$&&&$1$&\\
$O_{Dt}$&&$E/v$&$E/v$&\\
$O_{tB\Phi}$&&&$E/v$&$E/v$\\
$O_{tW\Phi}$&&$E/v$&$E/v$&$E/v$\\
$O_{tB}$&&&$E^2/v^2$&$E^2/v^2$\\
\hline
\hline
\end{tabular}
\end{center}
\caption[]{The energy-dependence of dimension-six operators
in Eq.~(\ref{operators}) for couplings
$Zb\bb$, $Wt\bb$, $Zt\tb$ and $\gamma t\tb$. An overall
normalization ${v^2}/{\Lambda^2}$ has been factored out.
}
\label{one}
\end{table}


Now we present the experimental constraints on various operators.
 The most direct bounds on these operators come from the
measurement of the observables $R_b$ and $A^b_{FB}$ at LEP.
Updating the bounds in our previous paper Ref.~\cite{tth}
and assuming no accidental cancellation happens
between different operators, as is often assumed,
we give the limits below on each of operators in Eq.~(\ref{operators})
 at the $1\sigma$ ($3\sigma$) level as
\begin{eqnarray}\label{bound2}
-1\times 10^{-2}~ (-2\times 10^{-2})&<
\frac{v^2}{\Lambda^2}C_{qW}
<&-1\times 10^{-4}~ (1\times 10^{-2}), \nonumber \\
-2\times 10^{-2}~ (-5\times 10^{-2})&<
\frac{v^2}{\Lambda^2}C_{qB}
<&-3\times 10^{-4}~ (2\times 10^{-2}), \nonumber \\
5\times 10^{-5}~ (-4\times 10^{-3})&<
\frac{v^2}{\Lambda^2}C^{(1)}_{\Phi q}
<&4\times 10^{-3}~ (8\times 10^{-3}), \nonumber \\
5\times 10^{-5}~ (-4\times 10^{-3})&<
\frac{v^2}{\Lambda^2}C^{(3)}_{\Phi q}
<&4\times 10^{-3}~ (8\times 10^{-3}),
\end{eqnarray}
where $v=246\ {\gev}$ is the vacuum expectation
value of the Higgs field.
 One can see that the constraints on some of the operators
listed in Table~\ref{one} are relatively poor and there is
room for possible new physics.
However if the operators are not independent for a given model,
cancellations may happen among different contributions, therefore
the bounds obtained from $R_b$ and $A^b_{FB}$
may not be as restrictive \cite{santiago}.

Operators $O_{t2}$, $O_{Dt}$, $O_{tW\Phi}$, $O_{tB\Phi}$ and $O_{tB}$,
are not constrained by $R_b$ at tree level.
However, at one-loop level they contribute to gauge
boson self-energies. The authors of Ref.~\cite{Renard} have
considered these constraints and showed
some rather loose bounds on them. One can also put limits on various
coefficients of the operators using
the argument of partial wave unitarity. The upper bounds
are obtained for $\Lambda \approx 3-1$ TeV in Ref.~\cite{Renard,tth}
\begin{eqnarray}\label{bound3}
&& |C_{t2}|{v^2\over \Lambda^2} \simeq 0.29 - 2.6, \nonumber\\
&& |C_{Dt}|{v^2\over \Lambda^2} \simeq 0.07-0.63  \ \ \mbox{or } \ \
|C_{Dt}|{v^2\over \Lambda^2} \simeq -(0.04-0.40), \nonumber\\[0.1cm]
&& |C_{tW\Phi}|{v^2\over \Lambda^2} \simeq 0.02 - 0.15\ ,~~~
|C_{tB\Phi}|{v^2\over \Lambda^2} \simeq 0.02 - 0.15\ , \nonumber \\
&& |C_{tB}|{v^2\over \Lambda^2} \simeq 0.04-0.34  \ \ \mbox{or } \ \
|C_{tB}|{v^2\over \Lambda^2} \simeq -(0.03-0.29).
\end{eqnarray}
Bounds on $C_{Db}$ and $C_{bW\Phi}$ are very weak due to their
small contributions to $Z\to b\bar b$ decay.

\section{Top-quark spin correlation at linear colliders
with anomalous couplings}

We study in this section the production and decay of top-quark
pair in the presence of anomalous couplings
and examine the different behavior of top-quark
spin correlations in various spin bases.

\subsection{The spin configuration in $t\bar t$ production}
We consider the top-quark pair production in $e^+e^-$ collisions
\begin{equation}
e^+e^- \to V^*\to t\bar t,\ \ \ V=\gamma, Z.
\end{equation}
To make our discussion general we write the
effective CP-even vertices of $Zt\tb$ and $\gamma t\tb$ as
\bea \label{vtt}
\Gamma_{Vt\tb}^{\mu}=ieQ_{tR}^{V}\gamma^{\mu}P_R
+ieQ_{tL}^{V}\gamma^{\mu}P_L+ieQ_{t\tb}^{V}\frac{(p_{t}-
p_{\tb})^{\mu}}{m_t},
\eea
where $P_{L,R}$ are the projection operators $(1\mp \gamma_5 )/2$ and
$p_t$ and $p_{\tb}$ are the momenta of the outgoing top quark and
top antiquark, respectively.
We have neglected the terms which
contain $k^{\mu}=(p_t+p_{\tb})^{\mu}$ by using the Dirac equation
for massless electrons.
In terms of the gauge-invariant operators in Eq.~(\ref{operators}),
the left(right)-handed couplings $Q_{tL}^V(Q_{tR}^V)$ and
the electroweak magnetic-moment couplings $Q_{t\tb}^V$ are
\bea \label{ctt}
Q_{tR}^{\gamma}&=&Q_t \left(1-\frac{C_{tB}}{\Lambda^2}\frac{c_W k^2}{e Q_t}
+\frac{C_{tW\phi}s_W+C_{tB\phi}c_W}{\Lambda^2}\frac{8}{\sqrt{2}}
\frac{m_W m_t}{geQ_t}\right), \nonumber\\
Q_{tL}^{\gamma}&=&Q_t \left(1-
\frac{C_{qW}s_W+C_{qB}c_W}{\Lambda^2}\frac{k^2}{e Q_t}
+\frac{C_{tW\phi}s_W+C_{tB\phi}c_W}{\Lambda^2}\frac{8}{\sqrt{2}}
\frac{m_W m_t}{geQ_t}\right), \nonumber\\
Q_{tR}^Z&=&Q_t^R \left(1+\frac{C_{tB}}{\Lambda^2}\frac{s_W k^2}{e Q_t^R}
-\frac{C_{t2}}{\Lambda^2} \frac{m_Z v}{e Q_t^R}
+\frac{C_{tW\phi}c_W-C_{tB\phi}s_W}{\Lambda^2}\frac{8}{\sqrt{2}}
\frac{m_W m_t}{geQ_t^R}\right), \nonumber\\
Q_{tL}^Z&=&Q_t^L (1-
\frac{C_{qW}c_W-C_{qB}s_W}{\Lambda^2}\frac{k^2}{e Q_t^L}
-\frac{C_{\phi q}^{(1)}-C_{\phi q}^{(3)}}{\Lambda^2}
\frac{m_Z v}{e Q_t^L} \nonumber\\&&
+\frac{C_{tW\phi}c_W-C_{tB\phi}s_W}{\Lambda^2}\frac{8}{\sqrt{2}}
\frac{m_W m_t}{geQ_t^L}), \nonumber\\
Q_{t\tb}^{\gamma}&=&-\frac{C_{tW\phi}s_W+C_{tB\phi}c_W}{\Lambda^2}
\frac{4}{\sqrt{2}}\frac{m_W m_t}{ge},\nonumber\\
Q_{t\tb}^{Z}&=&-\frac{C_{tW\phi}c_W-C_{tB\phi}s_W}{\Lambda^2}
\frac{4}{\sqrt{2}}\frac{m_W m_t}{ge}
+\frac{C_{Dt}}{\Lambda^2}\frac{m_Z m_t}{2\sqrt{2}e},
\eea
 where
$Q_e=-1$, $Q_t=\frac{2}{3}$,
$Q_{eR}=\frac{s_W}{c_W}$, $Q_{eL}=\frac{2s_W^2-1}{2s_W c_W}$,
$Q_t^R=-\frac{2s_W}{3c_W}$ and $Q_t^L=\frac{3-4s_W^2}{6s_W c_W}$,
which are the SM couplings.

\begin{figure}[htb]
\centerline{\epsfysize 2 truein \epsfbox{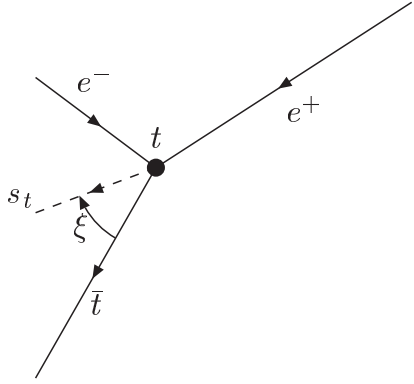}}
\caption{The spin axis $\vec s_t$ defined in the top-quark
rest-frame with respect to the anti-top momentum direction.}
\end{figure}
The discussion on the polarized top-quark production requires
a definite spin basis. In the SM, it is shown that the size of
top-spin correlation for different spin bases are quite distinct
\cite{parke,mahlon,ms}.
In the presence of anomalous couplings, we adopt the same notation
of generic spin basis~\cite{parke}. A general spin axis $\vec s_t$
for the top quark
is defined by an angle $\xi$ (clockwise) in the $t$ rest-frame
with respect to $\vec p_{\bar t}$ direction, as shown in Fig.~1.
The top antiquark spin axis can be defined with the same angle $\xi$,
but in the $\bar t$ rest-frame with respect to $\vec p_{t}$
direction. For instance, a polarized state $t_\uparrow \tb_\uparrow$
refers to a top quark with spin along $+\vec s_t$ and an anti-top
quark with spin along $+\vec s_{\bar t}$.
With this choice of spin basis,
the tree level differential polarized cross sections are given
in terms of the center-of-mass energy $\sqrt{s}$
and top-quark speed $\beta=\sqrt{1-4m_t^2/s}$ as
\bea\label{eett}
{d\sigma(e_L^- e_R^+ \to t_\uparrow \tb_\uparrow)\over d \cos\theta}
&=&
{d\sigma(e_L^- e_R^+ \to t_\downarrow \tb_\downarrow)\over d \cos\theta}
\nonumber\\
&=&
(\frac{3\pi\alpha^2}{2s}\beta)
|A_{LR} \cos \xi-B_{LR}\sin\xi
-2f_{t\tb}^L\gamma\beta^2 \sin\theta \cos\xi|^2,
\nonumber\\
{d\sigma(e_L^- e_R^+ \to t_\uparrow \tb_\downarrow)\over d \cos\theta}
&=&
{d\sigma(e_L^- e_R^+ \to t_\downarrow \tb_\uparrow)\over d \cos\theta}
\nonumber\\
&=&
(\frac{3\pi\alpha^2}{2s}\beta)
|A_{LR} \sin \xi+B_{LR}\cos \xi \pm D_{LR}
-2f_{t\tb}^L\gamma\beta^2 \sin\theta \sin\xi|^2,
\eea
where $\theta$ is the top scattering angle with respect to the $e^-$
beam and $\gamma=1/\sqrt{1-\beta^2}$, and
\bea\label{abd}
A_{LR}&=&[(f_{LL}+f_{LR})\sqrt{1-\beta^2} \sin\theta]/2,\nonumber\\
B_{LR}&=&[f_{LL}(\cos\theta + \beta) + f_{LR}(\cos\theta - \beta)]/2,
\nonumber\\
D_{LR}&=&[f_{LL}(1+\beta \cos \theta) + f_{LR}(1-\beta \cos \theta)]/2,
\eea
with
\bea \label{ftt}
f_{IJ}&=&Q_e Q_{tJ}^{\gamma}+Q_{eI} Q_{tJ}^Z \frac{s}{s-m_Z^2+im_Z\Gamma_Z},
~~~~~I,J=R,L,
\nonumber\\
f_{t\tb}^I&=&Q_e Q_{t\tb}^{\gamma}+Q_{eI} Q_{t\tb}^Z
\frac{s}{s-m_Z^2+im_Z\Gamma_Z},
~~~~~I=R,L.
\eea
Similarly, one can obtain the differential cross sections
for a different beam polarization $e_R^- e_L^+$ by interchanging
the labels $R \leftrightarrow L$ and
$\uparrow\ \leftrightarrow\ \downarrow$ in Eq.~(\ref{eett}).

There are three typical bases charaterized by choosing a different
angle $\xi$ as follows.
\begin{itemize}
\item Helicity basis:
For $\cos \xi =+1\ (-1)$,
the top-quark spin axis is against (along) its direction of motion.

\item Beamline basis:
For $\cos \xi = (\cos\theta + \beta)/(1+\beta \cos \theta)$,
the top-quark spin is in the positron direction in the top rest-frame.

\item Off-diagonal basis: $\tan \xi = A_{LR}^{SM}/B_{LR}^{SM}$,
where $A_{LR}^{SM}$ and $B_{LR}^{SM}$ represent the SM parts of
$A_{LR}$ and $B_{LR}$ respectively in Eq.~(\ref{abd}).
Although one can similarly define another off-diagonal basis
for $e_R^- e_L^+$ scattering, the considerably small difference
between the two bases allows us to
use the off-diagonal basis for $e_L^- e_R^+$
even when discussing $e_R^- e_L^+$ scattering~\cite{parke}.
\end{itemize}
Regarding the off-diagonal basis, Parke and collaborators
have applied it to top-pair production at a linear
and hadron colliders \cite{parke,ms},
and found many advantages in studying physics of the top quark.
It gives rise to substantial spin correlation than the
helicity or beamline basis do~\cite{nasuno}. For example,
$\frac{d\sigma}{d \cos\theta}
(e_L^- e_R^+ \to t_\uparrow \tb_\uparrow)$ and
$\frac{d\sigma}{d \cos\theta}
(e_L^- e_R^+ \to t_\downarrow \tb_\downarrow)$
are equal to zero for off-diagonal basis at the SM tree level and therefore
all $t\tb$ pairs are of opposite spins.
Moreover, the dominant component up to 97\% of the $e^-_L e^+$ cross section
is $t_{\uparrow}\tb_{\downarrow}$ at the energy
$\sqrt{s}=400~\gev$, much larger than 58\% in the helicity basis~\cite{parke}.
Furthermore,
the top-pair production density matrix in the off-diagonal basis
is greatly simplified due to many zero entries in the matrix \cite{ms}.

In the presence of anomalous couplings, however,
the situation may be different due to the very different
helicity structures of the new operators.
Thus we will consider all of the three
bases in the following discussions in order to find out
which basis is more sensitive to which class of operators.
Of the eleven operators listed in Eq.~(\ref{operators}) we will
take two operators $O_{tB}$ and $O_{tW\phi}$ for a detail study.
We choose these two operators for the following reasons.
First, they have distinctive energy-dependence as seen
in Table~\ref{one}, which characterizes a typical
feature of anomalous couplings. Second, while the operator
$O_{t B}$ modifies $Zt {\bar t}$ and $\gamma t {\bar t}$
which affect the top-quark production, the operator $O_{tW\phi}$
modifies the top-quark decay as well.
In addition, these two operators have no direct effects on
$Z\to b \bb$ measurement, so their coefficients are not
strongly constrained. Consequently, the top-quark production
and decay would be the unique place to explore them if
they generate sizeable effects at high energy experiments.
Numerically taking the conservative limits in
Eq.~({\ref{bound3}}), we will adopt
\begin{equation}
|C_{tB}|{v^2\over \Lambda^2},\ \ |C_{tW\phi}|{v^2\over \Lambda^2}< 0.02,
\end{equation}
in our analysis.

\begin{figure}[thb]\label{c-s}
\centerline{\epsfysize 6.2 truein \epsfbox{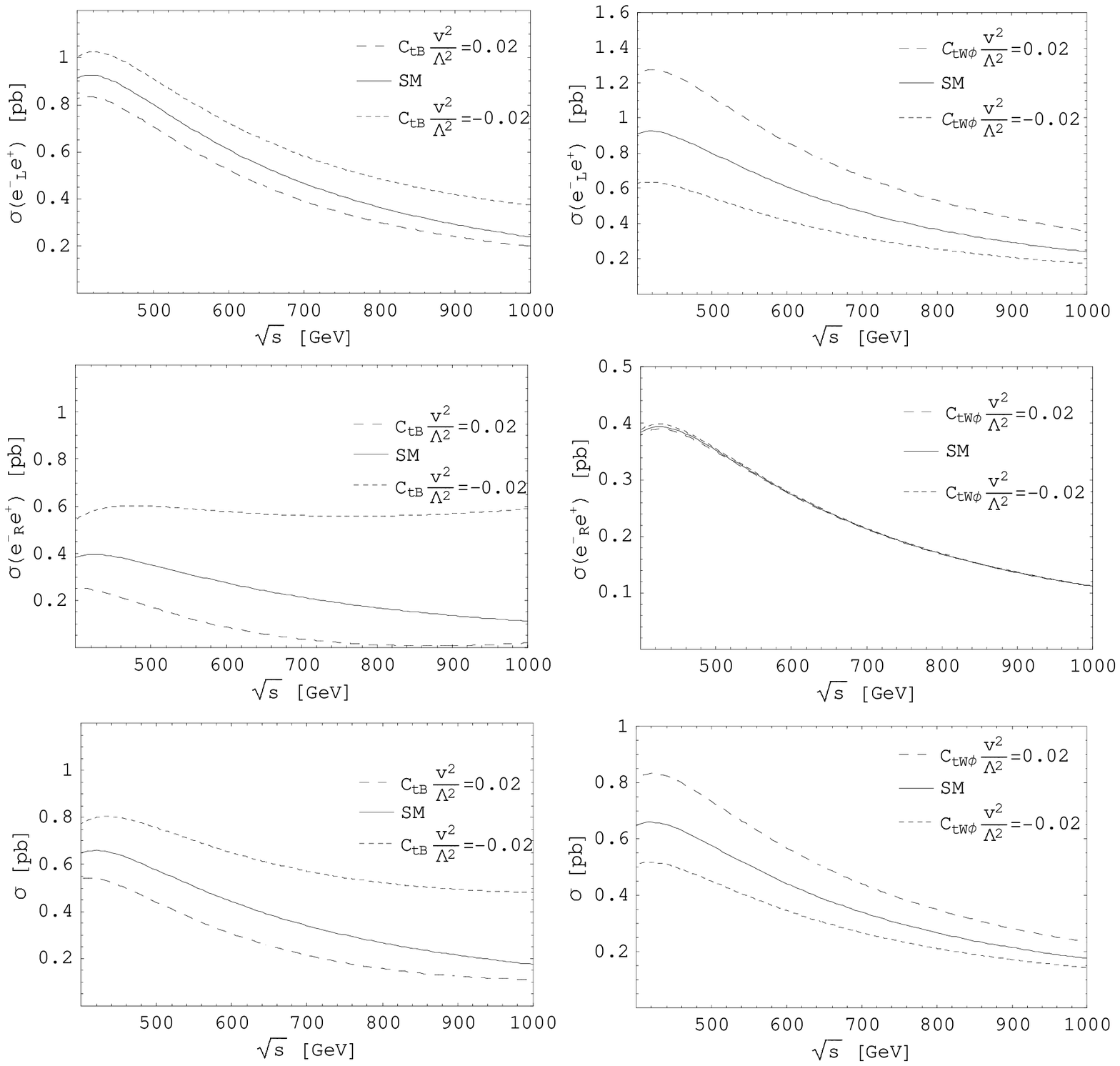}}
\caption[]{The total cross sections and polarized cross sections
for $e^+e^- \to t\bar t$ production
versus the $e^+e^-$ c.~m.~energy for
$O_{tB}$ and $O_{tW\phi}$. The solid curves are for the
SM prediction.}
\end{figure}

We plot in Fig.~2 the total cross section and various polarized cross sections
of $e^+e^- \to t\bar t$ versus $\sqrt s$ in the presence of operators
$O_{tB}$ and $O_{tW\phi}$. One can clearly see from
the figure the effects of new physics. However, the operator $O_{tW\phi}$
seems not to make any correction to $\sigma(e_R^-e^+)$. This is due to
the fact that the couplings of the right-handed electron with
$\gamma$ and $Z$ are from $U_Y(1)$ gauge interaction, while
$O_{tW\phi}$ are of purely $SU_L(2)$ interactions.
We have numerically verified the cancellation of
$O_{tW\phi}$ between $e^-_Re^+\to\gamma^*\to t\tb$ and
$e^-_Re^+\to Z ^*\to t\tb$ contributions.

In Fig.~3 and Fig.~4, we plot the energy dependence of the fraction of top
quark pair production in three different bases for
$O_{tB}$ and $O_{tW\phi}$, respectively.
The fraction is defined by $\sigma(e^-_{L/R} e^+ \to
t_{s_t} \tb_{s_{\bar t}})/\sigma_{total}$,
where $\sigma_{total}$ is the polarized total cross section
for the process $e^-_{L/R} e^+ \to t_{s_t} \tb_{s_{\bar t}}$
including appropriate anomalous couplings.
While the off-diagonal, beamline, and helicity bases present
the most, medium, and least polarization of $t\bar t$ states,
respectively, the percentage effects due to new operators
are of reversed order, namely it is the most significant
in the helicity basis instead.
The reason is as follows: The new physics effects come dominantly
from the interference terms between the SM and anomalous couplings
at the order of $1/\Lambda^2$, and the pure contributions of
the anomalous couplings set in at a higher order of $1/\Lambda^4$.
Since in the SM with the off-diagonal basis, the spin
configuration is almost exclusively $t_\uparrow \tb_\downarrow$,
the leading anomalous contribution coming from the interference
is forced to be in the same unique configuration.
On the other hand, the helicity basis presents a large
partition between $t_\uparrow \tb_\downarrow$ and $t_\downarrow
\tb_\uparrow$, so that it allows a more significant effect
in reshuffling these two configurations due to the existence
of the anomalous operators, thus leading to possibly appreciable
differences when comparing with the SM predictions.
We notice that because of the strong
energy dependence of the couplings,
the effects are generally larger at higher energies.
Moreover, we also observed a particularly large deviation
from the SM prediction for the operator $O_{tB}$ in
$e^-_R e^+_L$ reaction near $\sqrt s\approx 850$ GeV,
see Fig.~3.
This is due to the corrections from the direct term which
is proportion to $C^2_{tB}\times \frac{s^2}{\Lambda^4}$.
In the case the interference term gives the negative contributions and
the direct term gives the positive contributions, there should
be a peak at some energy scale and this scale is decided by
the size of $C_{tB}$ and $\Lambda$. For the parameter we
selected i.e. $C_{tB}v^2/\Lambda^2=0.02$, it appears at the
energy of 850 GeV.

In Fig.~5 and Fig.~6, we show the behavior of the fraction
at $\sqrt{s}=800~\gev$ as a function of the couplings
$O_{tB}$ and $O_{tW\phi}$, respectively. The SM prediction
can be read off from the figures for the couplings to be
zero. The sign of the coupling $O_{tB}$ may be inferred
by different beam polarizaton choices. For instance,
consider a $t_\uparrow \tb_\uparrow$ final state
as in Fig.~5, a positive coupling gives significant
enhancement to $e^-_L e^+_R$ initial state and a negative
coupling leads a reduction for $e^-_R e^+_L$ initial state.

Given the discussion above, we conclude that the
percentage corrections from anomalous couplings
to the spin configuration in the off-diagonal basis
are much smaller than that in the helicity basis for the polarized
$e^-e^+$ scattering.
The enhanced spin correlation in the SM is not necessarily
beneficial in distinguishing contributions from anomalous couplings.

\subsection{The effects on the correlation coefficient in top-quark decay}

We now consider the process of top decay and examine
the effects of anomalous couplings $Wtb$ on
the correlation coefficient.
In general the effective vertex of $Wtb$ coupling can be written as
\bea\label{coup}
\Gamma_{tWb}^{\mu}=i\frac{1}{\sqrt{2}}
[g_L\gamma^{\mu}P_L + g_R\gamma^{\mu}P_R
+ g_{tb}^L \frac{(p_t + p_b)^{\mu}}{m_t}P_L
+ g_{tb}^R \frac{(p_t + p_b)^{\mu}}{m_t}P_R],
\eea
where $p_t$ and $p_b$ are the momenta of incoming top quark and
outgoing bottom quark respectively and
\bea
g_L&=&gV_{tb}-\frac{C_{qW}}{\Lambda^2}(2k^2)
+\frac{C_{\phi q}^{(3)}}{\Lambda^2} \frac{4m_W^2}{g}
+\frac{C_{tW\phi}}{\Lambda^2} \frac{8m_Wm_t}{\sqrt{2}},
\nonumber\\
g_R&=&\frac{C_{bW\phi}}{\Lambda^2} \frac{8m_Wm_t}{\sqrt{2}g}
+\frac{C_{t3}}{\Lambda^2} \frac{2m_W^2}{g},
\nonumber\\
g_{tb}^L&=&-\frac{C_{bW\phi}}{\Lambda^2}\frac{8m_Wm_t}{\sqrt{2}g}
-\frac{C_{Db}}{\Lambda^2}\frac{m_Wm_t}{\sqrt{2}},
\nonumber\\
g_{tb}^R&=&-\frac{C_{tW\phi}}{\Lambda^2}\frac{8m_Wm_t}{\sqrt{2}g}
-\frac{C_{Dt}}{\Lambda^2}\frac{m_Wm_t}{\sqrt{2}}.
\eea
Note that the SM coupling is the first term in $g_L$.
It is easy to see from Eq.~(\ref{coup}) that the leading
corrections to the top decay at order $1/\Lambda^2$ come
from interference between the SM and terms in $g_L$ and
terms in $g_{tb}^R$. We will henceforth neglect effects
from $g_R$ and $g_{tb}^L$ which are suppressed by
$1/\Lambda^4$.

Using the narrow width approximation for the $W$ boson
\bea
\frac{1}{ (k^2-m^2)^2+m^2\Gamma^2 } \to
\frac{\pi}{m\Gamma}\delta(k^2-m^2),
\eea
we express the polarized differential top-decay width
for $t\to b f\bar f'$ as
\bea\label{width}
\frac{d \Gamma{(t_{\uparrow}~{\rm or}~t_{\downarrow})}}
{d\cos\theta_{D}}
=\frac{m_t}{16(4\pi)^2} \frac{g^2}{\bar y} \frac{\pi g_L^2}{\Gamma_W/m_W}
\frac{1}{6}(1-\bar y)^2(1+2\bar y)(1+\frac{2g_{tb}^R}{g_L}
\frac{1-\bar y}{1+2\bar y})(1\pm\cos\theta_{D})
                          \nonumber\\+{\cal O}(\frac{1}{\Lambda^4}),
\eea
where $\bar y=m_W^2/m_t^2$ and $\theta_{D}$ is the angle
between the momentum direction of the lepton or down-type
light quark and the top-spin axis in the top-quark rest-frame.
Here, we neglect all fermion mass except top quark.
The polarized differential decay rate in the SM is conveniently
parameterized as
\bea \label{rate}
\frac{1}{\Gamma} \frac{d \Gamma}{d\cos \theta_i}
=\frac{1\pm\alpha_i \cos\theta_i}{2}
\eea
where $i=b$, $D$ (lepton or down-type light quark) or $U$ (neutrino or
up-type light quark). The correlation coefficients $\alpha_i$
are $1, -0.31$ and $-0.41$ for $D$, $U$ and $b$, respectively~\cite{kuhn}.

Comparing Eq.~(\ref{width}) with Eq.~(\ref{rate}), we find that
the anomalous couplings do not give corrections to
the correlation coefficient $\alpha_{D}$ up to ${\cal O}(\frac{1}{\Lambda^2})$.
Since all spin bases are defined in the process of top production,
the angular distributions of different decay products
give same relationship of different spin bases.
Therefore we select the angle distribution between the momentum
direction of the lepton or down-type light quark and the top-spin axis
at top rest frame, whose correlation coefficient $\alpha_{D}$ is lagest.
We will show in the next subsection that this selection is advantageous
for us to define observables with large value.

We have also recalculated the top-decay width in the presence of the anomalous
coupling, which is shown in Table~\ref{two}. We find the corrections
may be as large as $5\%$. It would be very interesting to determine
the top width or individual branching fractions to this level
of accuracy.

\begin{table}[tb]\centering
\begin{tabular}{ll}
\cline{1-2}
\vbox to1.88ex{\vspace{1pt}\vfil\hbox to18.80ex{\hfil \hfil}} &
\vbox to1.88ex{\vspace{1pt}\vfil\hbox to15.60ex{\hfil $\Gamma$ [GeV]\hfil}} \\

\cline{1-2}
\vbox to1.88ex{\vspace{1pt}\vfil\hbox to18.80ex{\hfil SM ($C_i=0$)\hfil}} &
\vbox to1.88ex{\vspace{1pt}\vfil\hbox to15.60ex{\hfil 0.178\hfil}} \\

\cline{1-2}
\vbox to1.88ex{\vspace{1pt}\vfil\hbox to18.80ex{\hfil
$C_{qW}v^2/\Lambda^2=0.02$\hfil}} &
\vbox to1.88ex{\vspace{1pt}\vfil\hbox to15.60ex{\hfil 0.177\hfil}} \\

\cline{1-2}
\vbox to1.88ex{\vspace{1pt}\vfil\hbox to18.80ex{\hfil
$C_{qW}v^2/\Lambda^2=-0.02$\hfil}} &
\vbox to1.88ex{\vspace{1pt}\vfil\hbox to15.60ex{\hfil 0.180\hfil}} \\

\cline{1-2}
\vbox to1.88ex{\vspace{1pt}\vfil\hbox to18.80ex{\hfil
$C_{tW\phi}v^2/\Lambda^2=0.02$\hfil}} &
\vbox to1.88ex{\vspace{1pt}\vfil\hbox to15.60ex{\hfil 0.188\hfil}} \\

\cline{1-2}
\vbox to1.88ex{\vspace{1pt}\vfil\hbox to18.80ex{\hfil
$C_{tW\phi}v^2/\Lambda^2=-0.02$\hfil}} &
\vbox to1.88ex{\vspace{1pt}\vfil\hbox to15.60ex{\hfil 0.169\hfil}} \\

\cline{1-2}
\vbox to1.88ex{\vspace{1pt}\vfil\hbox to18.80ex{\hfil
$C_{Dt}v^2/\Lambda^2=0.02$\hfil}} &
\vbox to1.88ex{\vspace{1pt}\vfil\hbox to15.60ex{\hfil 0.177\hfil}} \\

\cline{1-2}
\vbox to1.88ex{\vspace{1pt}\vfil\hbox to18.80ex{\hfil
$C_{Dt}v^2/\Lambda^2=-0.02$\hfil}} &
\vbox to1.88ex{\vspace{1pt}\vfil\hbox to15.60ex{\hfil 0.179\hfil}} \\

\cline{1-2}
\end{tabular}
\caption[]{The leptonic decay width ($t \to W^+ b \to \bar{l} \nu b$)
containing the contributions from the SM plus anomalous couplings.}
\label{two}
\end{table}

\subsection{Spin asymmetry and spin-spin asymmetry}

Since we need the construction of the full event kinematics
for $t\bar t$ decays, the dileptonic decay mode would
be difficult to make use of due to the two missing neutrinos.
Although the identification of six-jet events in the pure
hardonic decay mode should be viable, it would be impossible
to sort out quarks from anti-quarks in the four light jets.
The semileptonic mode becomes the unique channel for the
study of top-spin correlations.
We consider the case where one lepton with positive
charge comes from
top-quark decay while two partons come from top antiquark decay.
To obtain the differential cross section of
$e^-e^+ \to t\tb \to (b \bar l \nu_{\bar l})(\bar b U D)$,
the production and decay spin density matrix are always used as
the usual approach in the approximation of narrow widths for top quark
and $W$ boson~\cite{choi,kuhn}. After integrating over all azimuthal
angles, one gets the double differential rate as
\bea
\frac{1}{\sigma}\frac{d^2\sigma}{d(\cos\theta_{\bar l})d(\cos\theta_{D})}
=\frac{1}{4}
(1+\langle s_t\rangle \alpha_{\bar l}\cos\theta_{\bar l}
+\langle s_{\tb}\rangle\alpha_{D}\cos\theta_{D}
+\langle s_t s_{\tb}\rangle \alpha_{\bar l}\alpha_{D}
\cos\theta_{\bar l}\cos\theta_{D})
\nonumber
\eea
where $\bar l$ ($D$) is the decay product of top quark (top antiquark)
and  $\cos\theta_{\bar l}$ ($\cos\theta_{D}$)
is the angle defined in Eq~(\ref{width}).
$\langle s_t\rangle$, $\langle s_{\tb}\rangle$
and $\langle s_t s_{\tb}\rangle$ are $t$, $\tb$ and $t\tb$
correlation functions, respectively. They can be expressed as
\bea
\langle s_t\rangle &=&
\sum_{I,J=L,R}
(R(e_I^- e_J^+ \to t_\uparrow \tb_\uparrow)
+R(e_I^- e_J^+ \to t_\uparrow \tb_\downarrow)
-R(e_I^- e_J^+ \to t_\downarrow \tb_\uparrow)
-R(e_I^- e_J^+ \to t_\downarrow \tb_\downarrow)),
\nonumber\\
\langle s_{\tb}\rangle&=&
\sum_{I,J=L,R}
(R(e_I^- e_J^+ \to t_\uparrow \tb_\uparrow)
-R(e_I^- e_J^+ \to t_\uparrow \tb_\downarrow)
+R(e_I^- e_J^+ \to t_\downarrow \tb_\uparrow)
-R(e_I^- e_J^+ \to t_\downarrow \tb_\downarrow)),
\nonumber\\
\langle s_t s_{\tb}\rangle &=&
\sum_{I,J=L,R}
(R(e_I^- e_J^+ \to t_\uparrow \tb_\uparrow)
-R(e_I^- e_J^+ \to t_\uparrow \tb_\downarrow)
-R(e_I^- e_J^+ \to t_\downarrow \tb_\uparrow)
+R(e_I^- e_J^+ \to t_\downarrow \tb_\downarrow)).
\nonumber
\eea
where
\bea
R(e_I^- e_J^+ \to t_{s_{t}} \tb_{s_{\tb}})
=\frac{\sigma(e_I^- e_J^+ \to t_{s_{t}} \tb_{s_{\tb}})}
{\sigma_{T}},~~~~~~I,J=L,R ~{\rm and} ~s_{t},s_{\tb}=\uparrow,\downarrow.
\eea
Here, $\sigma(e_I^- e_J^+ \to t_{s_{t}} \tb_{s_{\tb}})$ are the
polarized cross section in top-pair production
which could be obtained by integrating
over $\cos\theta$ in Eq.~(\ref{eett}).
$\sigma_{T}$ is the total cross section of $e^-e^+\to t \tb$.

Integrating over $\cos\theta_{D}$, one obtains
\bea
\frac{1}{\sigma}
\frac{d\sigma}{d(\cos\theta_{\bar l})}
=
\frac{1}{2}(1+
\alpha_{\bar l}^{s}
\cos\theta_{\bar l}),
\eea
where the effective correlation  coefficient
$\alpha_{\bar l}^{s}$ is given by
\bea\label{ls}
\alpha_{\bar l}^{s}=\langle s_t\rangle \alpha_{\bar l}.
\eea
Such effective spin correlation corresponds to
the top-quark spin asymmetry which is some observable
and can be derived as
\bea
A_s=
\frac{
N( {\cos\theta_{\bar l}>0})
-N( {\cos\theta_{\bar l}<0})}
{ N( {\cos\theta_{\bar l}>0})
+N( {\cos\theta_{\bar l}<0})}
=\frac {\alpha_{\bar l}^{s}}{2}.
\eea

Similarly,
we integrate over $\cos\theta_{D}$ in the region of
\bea
0<\cos\theta_{D}<1
\eea
and obtain
\bea
\left .
\frac{1}{\sigma}
\frac{d\sigma}{d(\cos\theta_{\bar l})}
\right |_
{\cos\theta_{D}>0}
\propto
\frac{1}{2}(1+
\alpha_{\bar l}^{ss}
\cos\theta_{\bar l}),
\eea
where
\bea\label{lss}
\alpha_{\bar l}^{ss}=
\frac{2\langle s_{\tb}\rangle + \langle s_t s_{\tb}\rangle \alpha_{D}}
{2+\alpha_{D}\langle s_t\rangle}
\alpha_{\bar l}.
\eea
This effective correlation coefficient corresponds to
the spin-spin asymmetry which is defined as
\bea
A_{ss}=
\frac{
N( {\cos\theta_{\bar l}>0}, {\cos\theta_{D}>0})
-N( {\cos\theta_{\bar l}<0}, {\cos\theta_{D}>0})}
{ N( {\cos\theta_{\bar l}>0}, {\cos\theta_{D}>0})
+N( {\cos\theta_{\bar l}<0}, {\cos\theta_{D}>0})}
=\frac {\alpha_{\bar l}^{ss}}{2}.
\eea
Even though distinction of the down-type quark
and the up-type quark in the final decay state is impossible,
we can at least hope to measure $A_{ss}$ by a statistical
method \cite{D0}.

Eqs.~(\ref{ls}) and (\ref{lss}) are generically valid for other
decay product. One can simply replace $\alpha_{\bar l}$ by
$\alpha_{i}$, where $i$ denotes any decay product.
Therefore, the observables $A_{s}$ and $A_{ss}$ are
proportional to the correlations coefficient $\alpha_{i}$
and we choose the largest value
$\alpha_{i}=\alpha_{\bar l}$.

The spin asymmetry $A_{s}$ and spin-spin asymmetry
$A_{ss}$  are shown versus $\sqrt{s}$ for $O_{tB}$
in the case of right-hand polarized electron beam in Fig.~7, and for
$O_{tW\phi}$ in the case of left-hand polarized electron beam in Fig.~8.
With the analysis in the last two subsections,
it is not surprising to see that
the corrections to $A_{s}$ and $A_{ss}$ are the most significant in
the helicity basis for $O_{tB}$ and $O_{tW\phi}$.

For the left-hand polarized electron beam, the top quark with up-spin
is the dominant configuration and for the right-hand polarized
electron beam the dominant one is that with down-spin.
Due to the cancellation between different polarized beams, there
is no unique basis which is optimal for searching new physics.
In Fig.~9 and Fig.~10, we show the spin asymmetry $A_{s}$
and spin-spin asymmetry $A_{ss}$ versus $\sqrt{s}$
for $O_{tB}$ and $O_{tW\phi}$ in the case of unpolarized beam.
The corrections to $A_{s}$ and $A_{ss}$ are the most significant in
the helicity basis for $O_{tB}$ and in the off-diagonal basis
for $O_{tW\phi}$, related to the chirality structure of the
operators.

\section{Summary and conclusion}

We took an effective Lagrangian approach to new physics
in the top-quark sector and discussed the anomalous-coupling
effects on the spin correlation in top-quark pair production and decay.
In the framework of the generic spin basis proposed by Parke and
collaborators, we performed detail calculations of spin correlations,
spin asymmetry and spin-spin asymmetry for three different bases.
We found that the helicity basis is the best for polarized beam,
and for unpolarized beam the optimal choice of the spin basis
depends on the specific form of the new physics.
Specifically the helicity basis will be the best choice in probing for
new physics associated with $U_Y(1)$ type operators such as
$O_{tB}$, while the off-diagonal basis appears to be more suitable
for $SU_L(2)$ type operators such as $O_{tW\phi}$.

\vskip 1cm
{\it Acknowledgments}:
We thank E.L. Berger, K. Hagiwara, S. Parke and J.-M. Yang
for discussions.
T. Huang, Z.-H. Lin and X. Zhang were supported in part by
the NSF of China. T. Han was supported in part by
a US DOE grant No.~DE-FG02-95ER40896 and by Wisconsin Alumni
Research Foundation.

\begin{figure}[thb] \label{ctb01}
\centerline{\epsfysize 6.0 truein \epsfbox{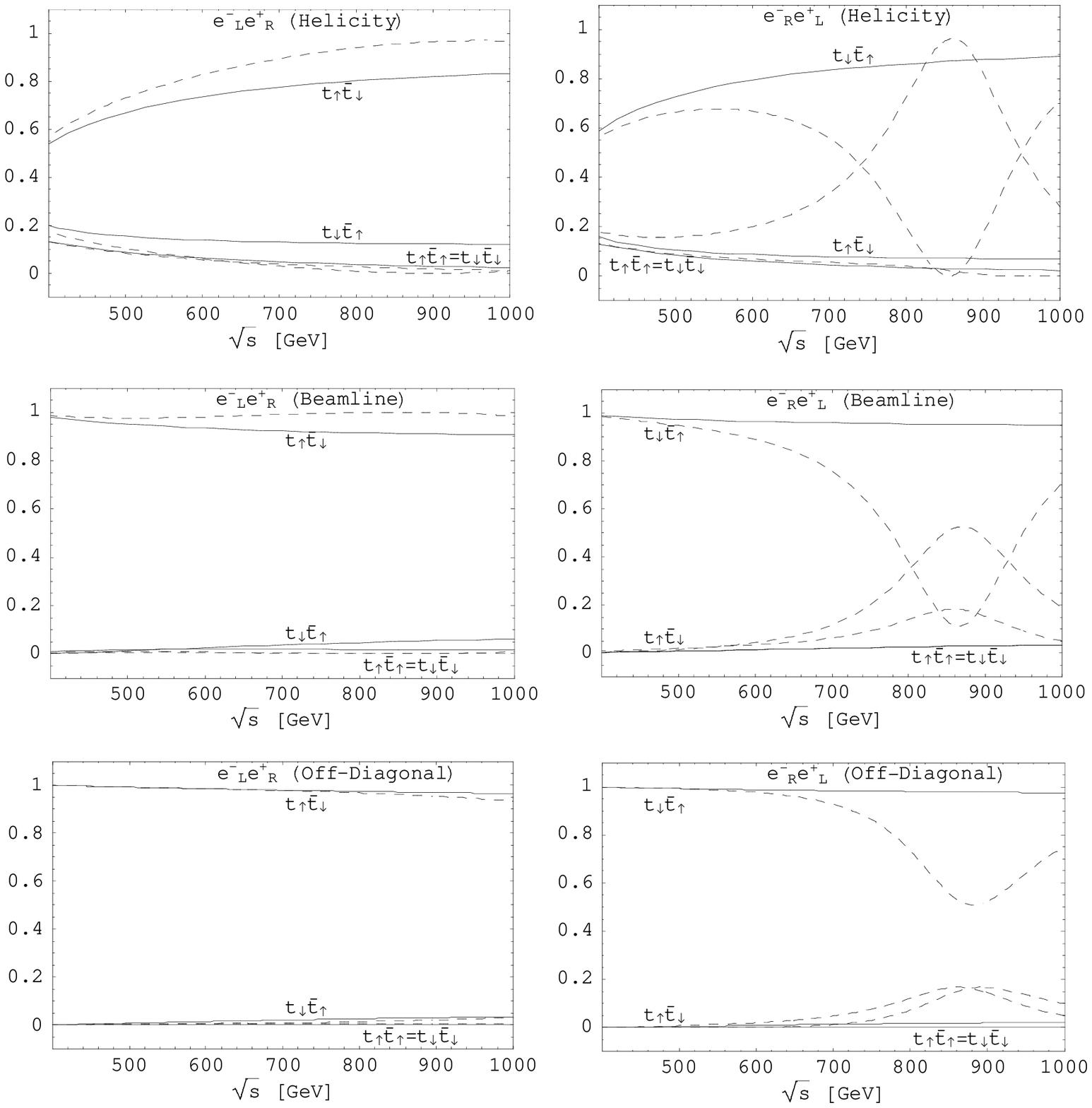}}
\caption[]{The fraction of top-quark pair production versus the
$e^+e^-$ c.~m.~energy
$\sqrt{s}$~[GeV]
for $O_{tB}$ in the
helicity, beamline and off-diagonal bases.
The solid curves are for the
SM expectation and the dashed curves are for $C_{tB}v^2/\Lambda^2=0.02$.}
\end{figure}

\begin{figure}[thb] \label{ctwp01}
\centerline{\epsfysize 6.0 truein \epsfbox{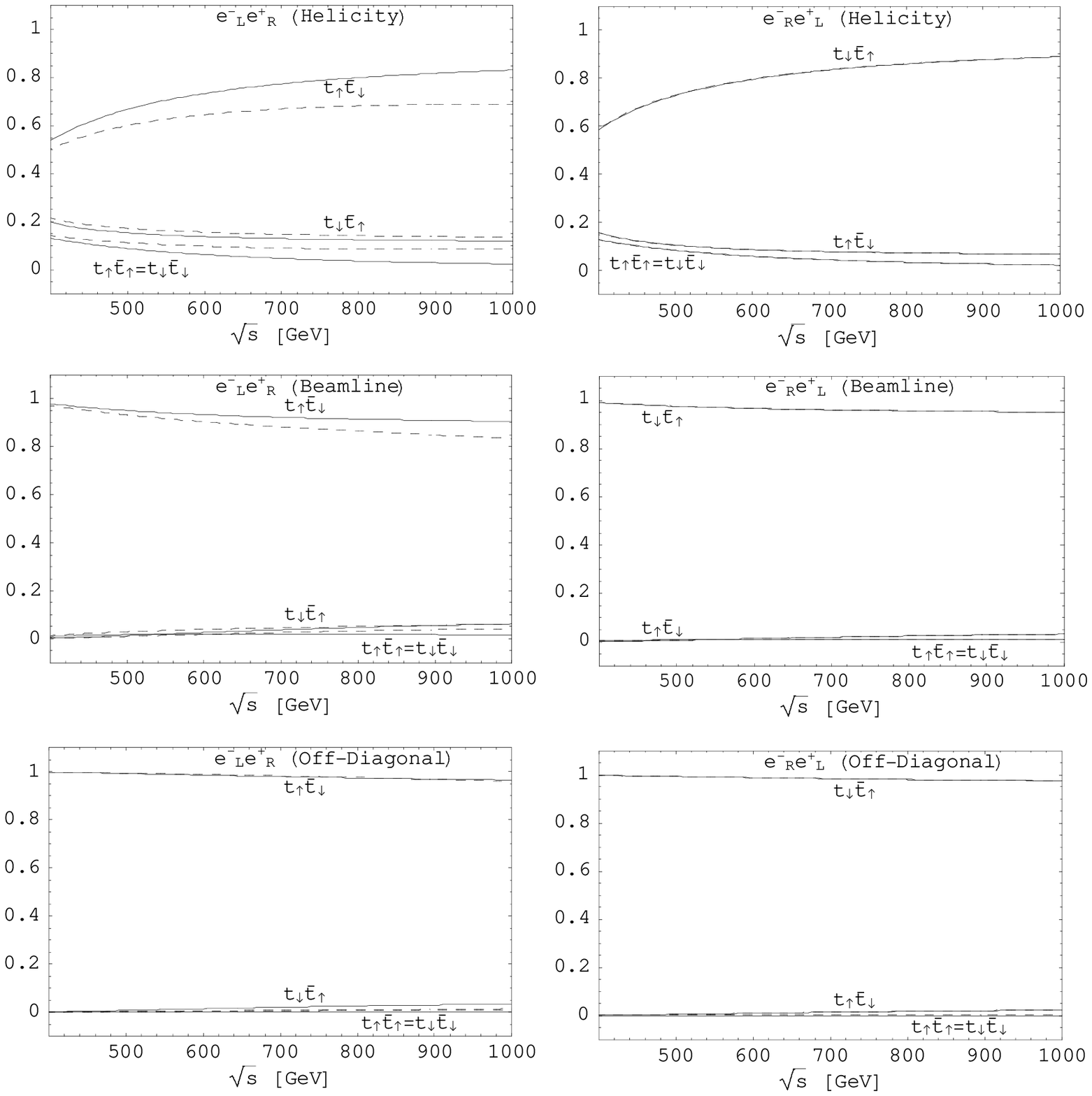}}
\caption[]{The fraction of top-quark pair production versus the
$e^+e^-$ c.~m.~energy $\sqrt{s}$~[GeV]
 for $O_{tW\phi}$ in the
helicity, beamline and off-diagonal bases.
The solid curves are for the
SM expectation and the dashed curves are for $C_{tW\phi}v^2/\Lambda^2=0.02$.}
\end{figure}

\begin{figure}[thb] \label{ctb02}
\centerline{\epsfysize 6.0 truein \epsfbox{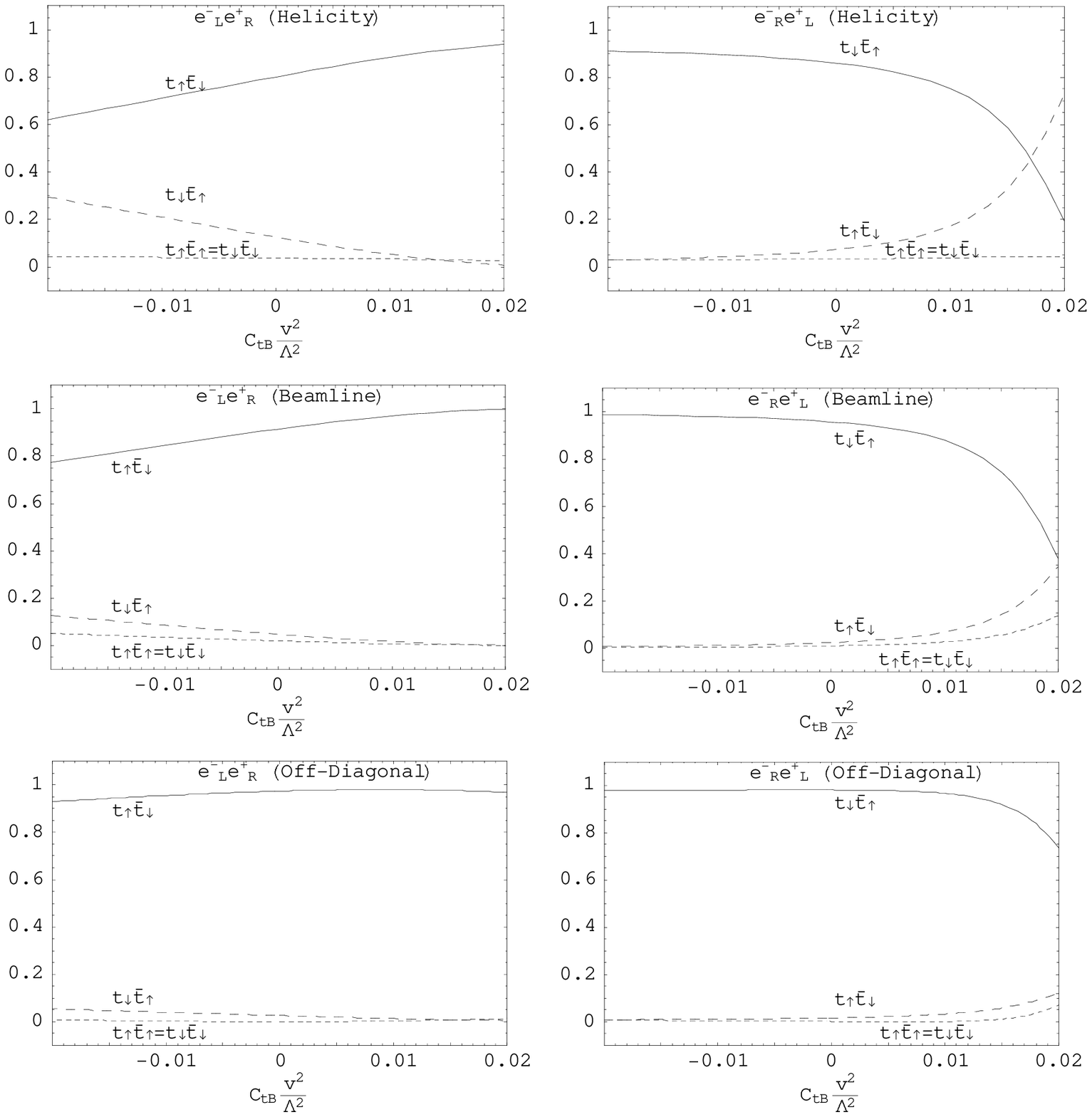}}
\caption[]{The fraction of top-quark pair production versus the couplings
for $O_{tB}$, $C_{tB}v^2/\Lambda^2$ and $\sqrt s=800$ GeV in the helicity,
beamline and off-diagonal bases .}
\end{figure}

\begin{figure}[thb]\label{ctwp02}
\centerline{\epsfysize 6.0 truein \epsfbox{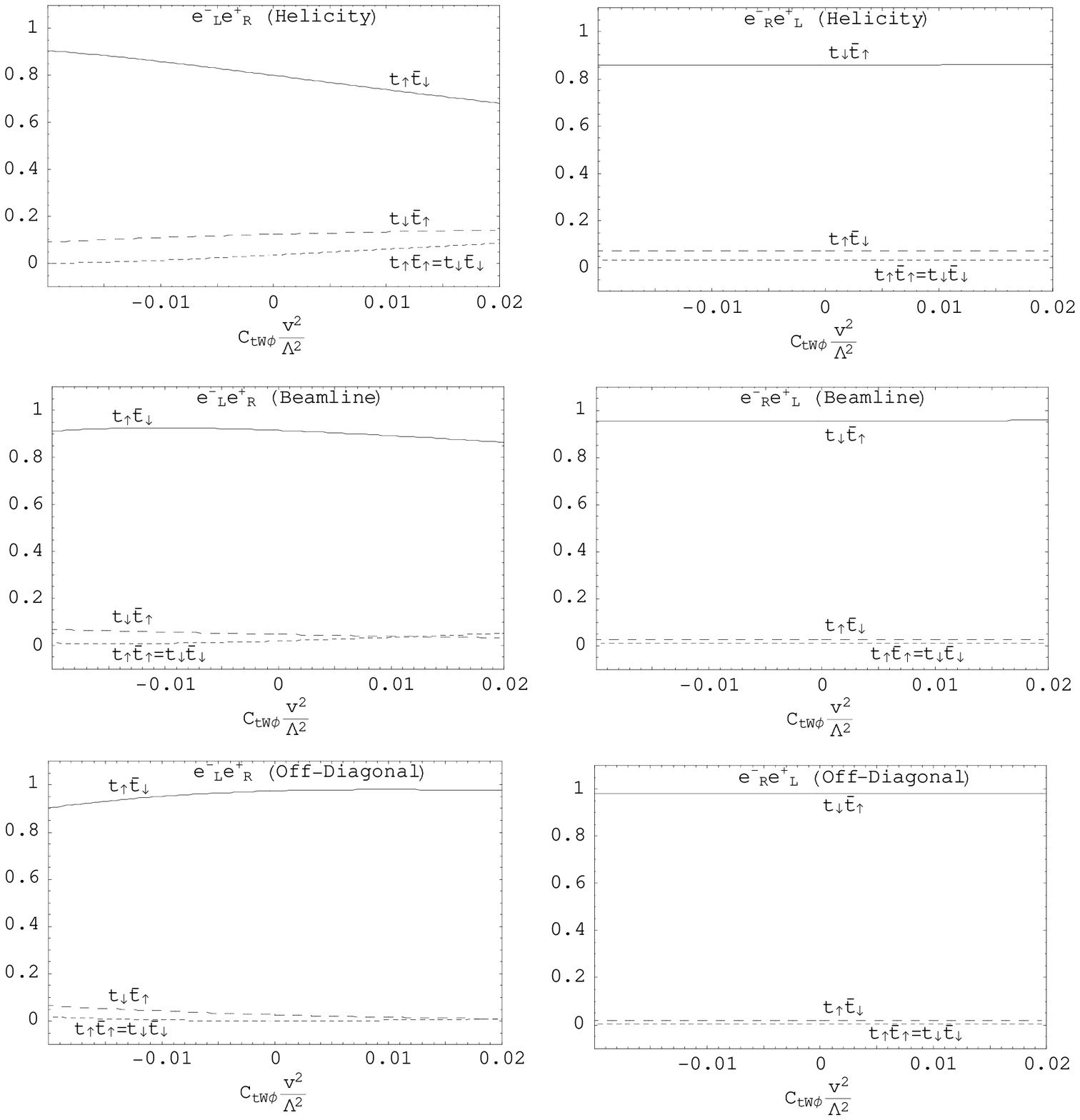}}
\caption[]{The fraction of top-quark pair production versus the couplings
for $O_{tW\phi}$, $C_{tW\phi}v^2/\Lambda^2$ and $\sqrt s=800$ GeV in the
 helicity, beamline and off-diagonal bases .}
\end{figure}

\begin{figure}[thb] \label{ctb03}
\centerline{\epsfysize 6.2 truein \epsfbox{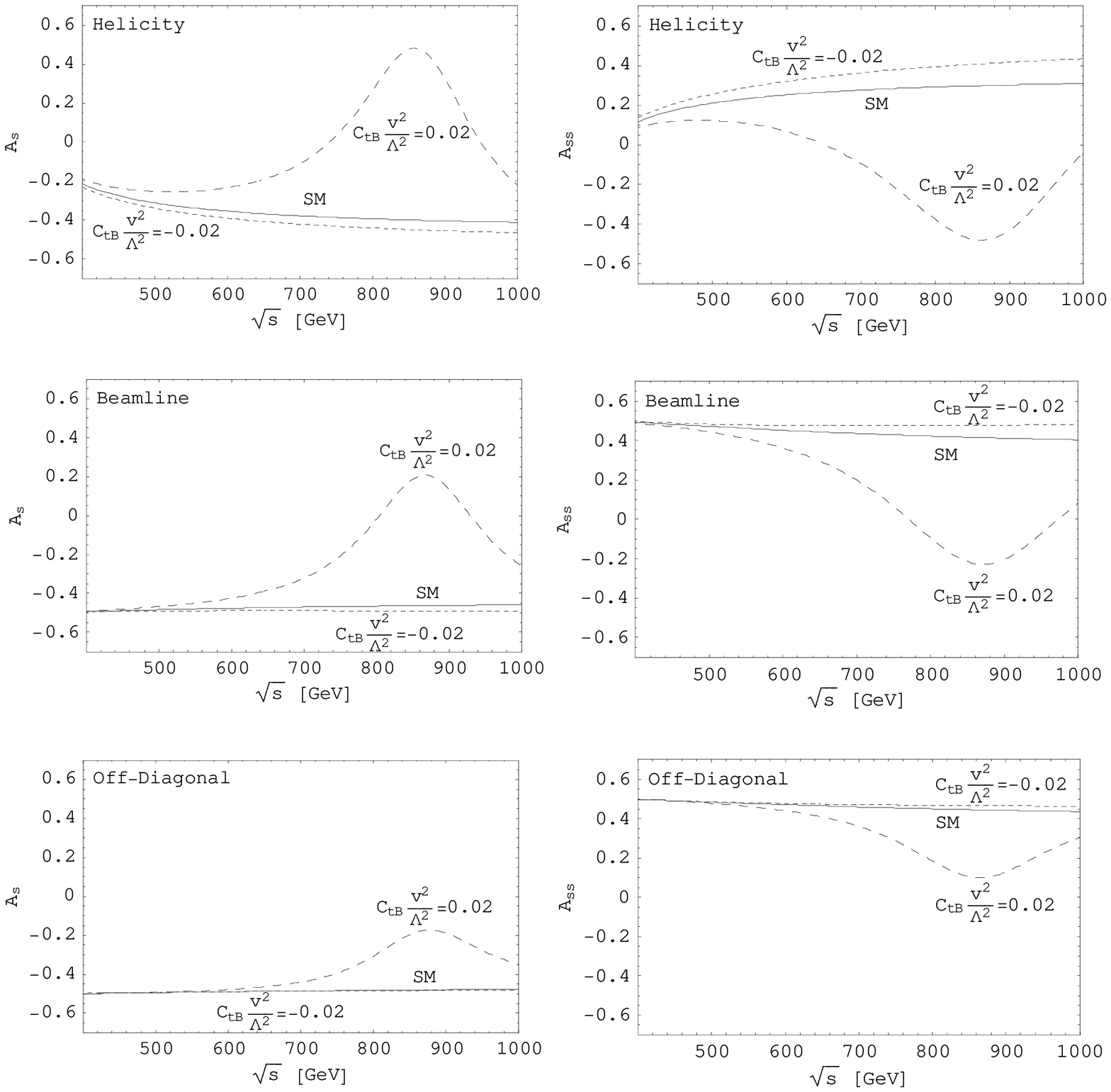}}
\caption[]{The spin asymmetry $A_{s}$ and spin-spin asymmetry $A_{ss}$
versus the $e^+e^-$ c.~m.~energy
for $O_{tB}$ in the helicity, beamline and off-diagonal bases
for right-hand polarized electron beam.}
\end{figure}

\begin{figure}[thb]\label{ctwp03}
\centerline{\epsfysize 6.2 truein \epsfbox{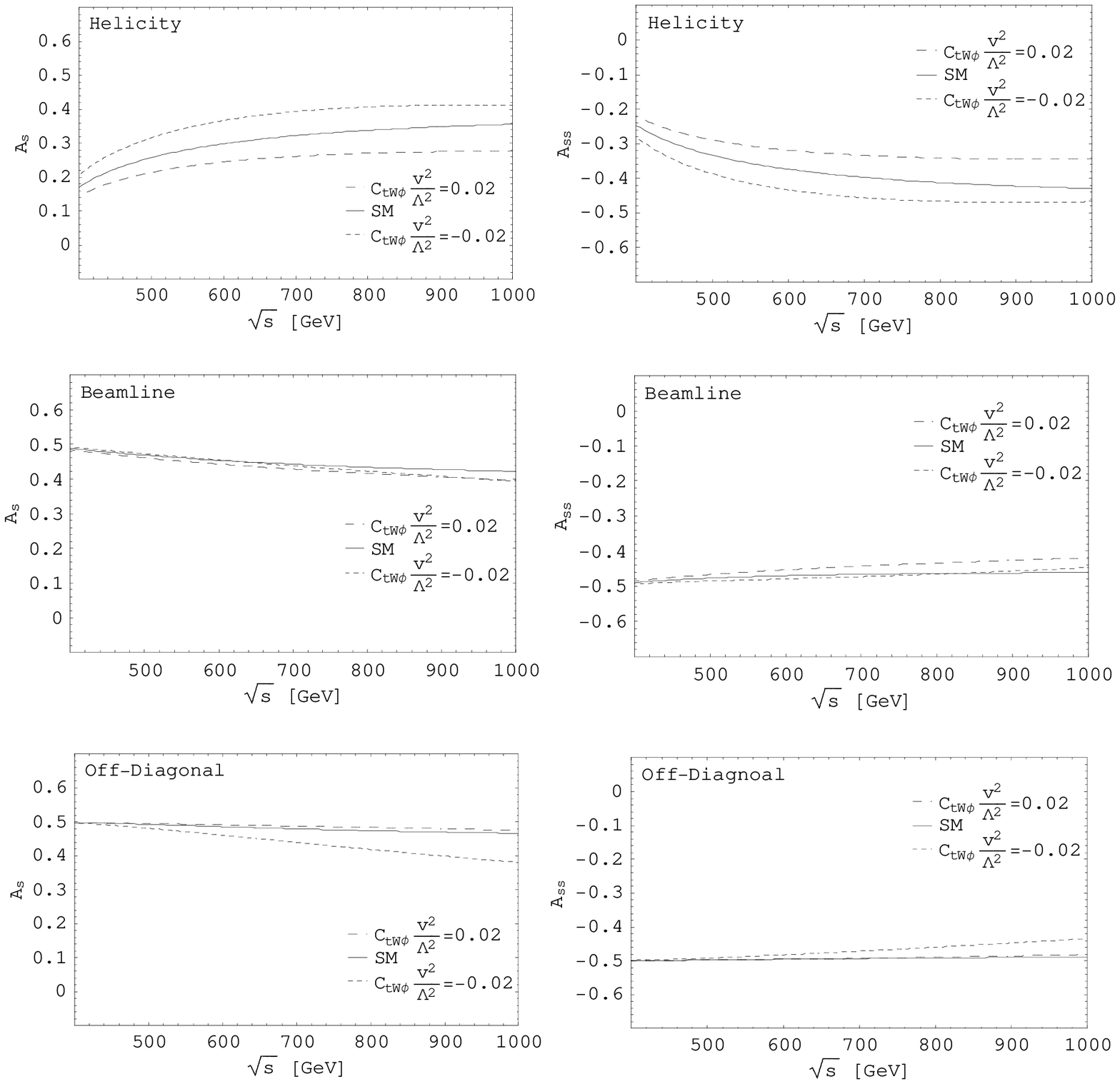}}
\caption[]{The spin asymmetry $A_{s}$ and spin-spin asymmetry $A_{ss}$
versus the $e^+e^-$ c.~m.~energy
for $O_{tW\phi}$ in the helicity, beamline and off-diagonal bases
for left-hand polarized electron beam.}
\end{figure}

\begin{figure}[thb] \label{ctb04}
\centerline{\epsfysize 6.2 truein \epsfbox{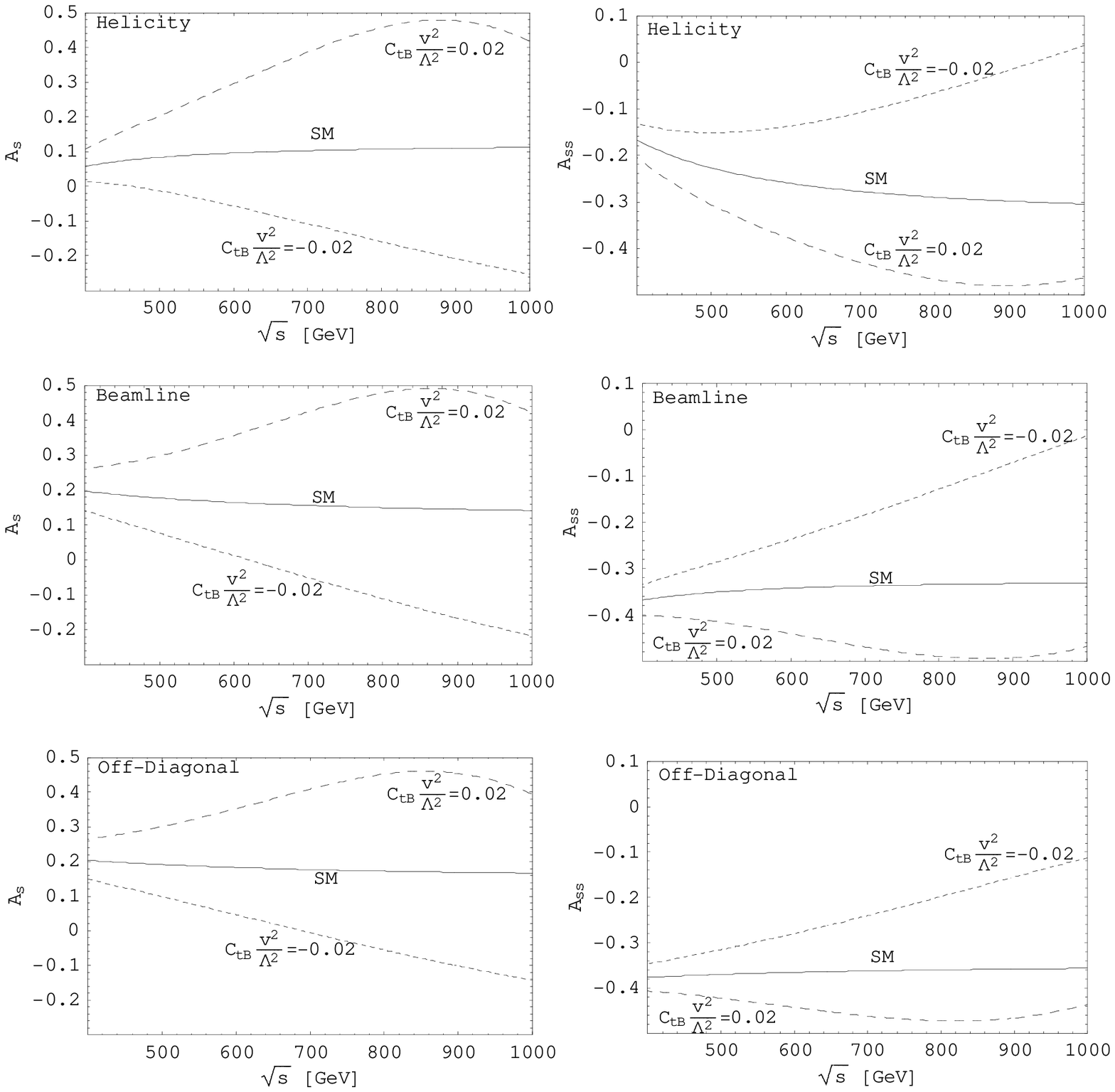}}
\caption[]{The spin asymmetry $A_{s}$ and spin-spin asymmetry $A_{ss}$
versus the $e^+e^-$ c.~m.~energy
for $O_{tB}$ in the helicity, beamline and off-diagonal bases
for unpolarized beam.}
\end{figure}

\begin{figure}[thb]\label{ctwp04}
\centerline{\epsfysize 6.2 truein \epsfbox{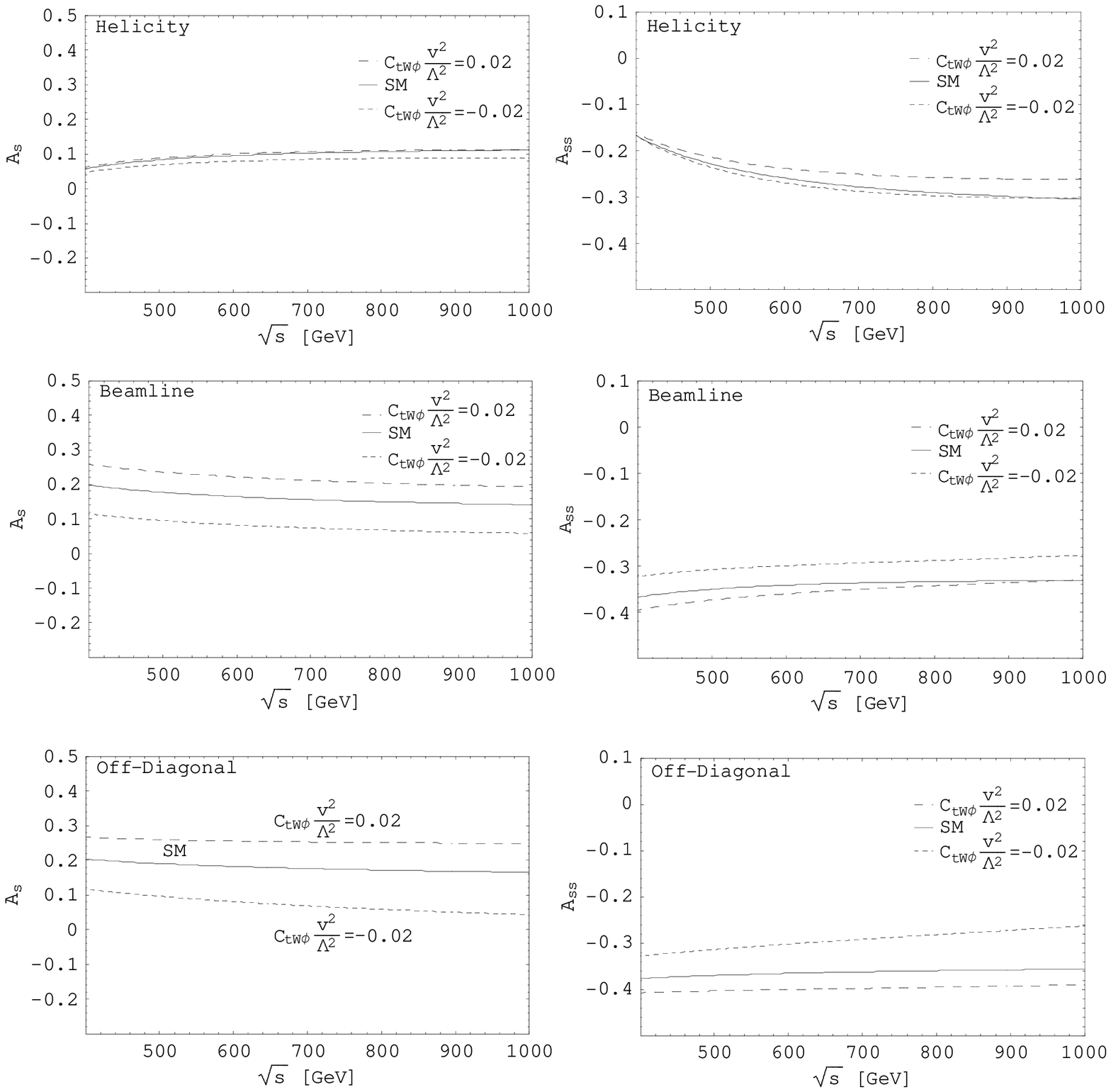}}
\caption[]{The spin asymmetry $A_{s}$ and spin-spin asymmetry $A_{ss}$
versus the $e^+e^-$ c.~m.~energy
for $O_{tW\phi}$ in the helicity, beamline and off-diagonal bases
for unpolarized beam.}
\end{figure}

\end{document}